\documentclass{ws-procs11x85}

\usepackage{balance}

\newcommand{\beq}{\begin{equation}}
\newcommand{\eeq}{\end{equation}}
\newcommand{\bea}{\begin{eqnarray}}
\newcommand{\eea}{\end{eqnarray}}
\newcommand{\gsim}{\lower.7ex\hbox{$\;\stackrel{\textstyle>}{\sim}\;$}}
\newcommand{\lsim}{\lower.7ex\hbox{$\;\stackrel{\textstyle<}{\sim}\;$}}
\makeindex
\begin{document}

\title{THEORETICAL PREDICTIONS FOR COLLIDER SEARCHES$^*$}

\author{G.~F. GIUDICE}

\address{Theoretical Physics Division, CERN, CH-1211 Geneva 23,
Switzerland}


\twocolumn[\maketitle\abstract{ I review recent developments in  
extensions of the Standard Model that address the question of
electroweak symmetry breaking and discuss how these theories 
can be tested at future colliders.}]

\footnotetext{$^*$Invited talk
presented at the 21st International Symposium on Lepton and Photon
Interactions at High Energies (Lepton Photon 2003), 11-16 August 2003,
Batavia, Illinois, USA.}

\baselineskip=13.07pt
In their search for understanding the fundamental laws of
elementary particles, high energy physicists are eager
to prove that one of the most successful and elegant scientific theories
ever formulated -- the Standard Model (SM) -- 
actually fails at distances
smaller than about a hundred zeptometers ($10^{-19}$~m). A generation of 
high energy colliders, culminating with the LHC under construction at CERN,
has been designed to achieve this goal. Here I will briefly review the
present status of the theoretical speculations for new physics within
reach of the LHC. In the spirit of this conference, and because of limited 
space, I will consider only topics
in which, in my opinion, there have been recent developments, leaving aside
other possibilities which, although interesting, have known only limited
progress. The list of references is also largely incomplete.

\section{``Big'' and ``Little'' Hierarchy Problems}\label{sec1}
The hierarchy problem\cite{hie} is not necessarily the most puzzling
open question of the SM, but it is certainly the most relevant to collider
experiments, since its resolution lies -- most likely -- in the TeV energy
range. Its formulation is well known. Treating the SM as an effective theory
valid up to a scale $\Lambda_{\rm SM}$, 
and cutting off momenta in loop integrals
at the same scale, we find that the dominant 
radiative correction to the Higgs mass is given by
\bea
\delta m_H^2&=&\frac{3G_F}{4\sqrt{2}\pi^2}\left( 2m_W^2+m_Z^2+m_H^2-4m_t^2
\right) \Lambda_{\rm SM}^2\nonumber \\
&=&-\left( 200~{\rm GeV}\frac{\Lambda_{\rm SM}}{0.7~
{\rm TeV}}\right)^2.
\eea
The request of no fine-tuning between the tree-level and one-loop contributions
to $m_H$ implies $\Lambda_{\rm SM}\lsim$~TeV. In other words, the SM cannot be 
valid beyond the TeV, and new physics should appear to modify the ultraviolet
behavior. I will refer to this result as the ``big'' 
hierarchy, since other fundamental scales,
such as the Planck mass $M_{\rm Pl}$, are known to be much larger than 
$\Lambda_{\rm SM}$.

As a word of caution, I should recall that, if I apply the same naturalness
argument
to the quartic divergences, then the present upper bound on the cosmological 
constant implies that the corresponding cut-off has to be smaller than 
$10^{-3}$~eV. Although we cannot fully rule out the possibility that the 
ultraviolet
behavior of gravity is prematurely modified at about $10^{-3}$~eV, 
this embarrassing result undoubtedly casts a grievous shadow over the
hierarchy problem.

Let me come back to the SM as an effective theory. Unknown new physics at the
cut-off is parametrized in terms of non-renormalizable operators. To be
most conservative, I will add only operators that preserve all local and global
SM symmetries and satisfy the criterion of minimal flavor 
violation\rlap.\,\cite{mfv}
As shown in Table~\ref{tab:1}, typical
limits on $\Lambda_{\rm LH}$, defined as the effective scale of the new
dimension-six operators (${\cal L}=\pm \Lambda_{\rm LH}^{-2}\cal O$), 
are $\Lambda_{\rm LH}>
5$--$10$~TeV.

\begin{table}
\begin{center}
\caption{90\% CL limits on the scale $\Lambda_{\rm LH}$ (in TeV)
of dimension-six
operators $\cal
O$ in the effective Lagrangian ${\cal L}=\pm \Lambda_{\rm LH}^{-2}\cal O$,
in both cases of constructive and destructive interference with the
SM contribution ($\pm$). The limits on the operators relevant to LEP1
are derived under the assumption of a light Higgs. 
\label{tab:1}} \vspace{0.2cm}
\begin{tabular}{c|c|c|c|}
&&$-$&$+$\\
\hline
\raisebox{0pt}[12pt][6pt]{LEP1\cite{lep1}} &
\raisebox{0pt}[12pt][6pt]{$H^\dagger \tau^a HW^a_{\mu\nu}B^{\mu\nu}$} &
\raisebox{0pt}[12pt][6pt]{10} &
\raisebox{0pt}[12pt][6pt]{9.7} \\
\raisebox{0pt}[12pt][6pt]{} &
\raisebox{0pt}[12pt][6pt]{$| H^\dagger D_\mu H|^2$} &
\raisebox{0pt}[12pt][6pt]{5.6} &
\raisebox{0pt}[12pt][6pt]{4.6} \\
\raisebox{0pt}[12pt][6pt]{} &
\raisebox{0pt}[12pt][6pt]{$iH^\dagger D_\mu H\bar L \gamma^\mu L$} &
\raisebox{0pt}[12pt][6pt]{9.2} &
\raisebox{0pt}[12pt][6pt]{7.3} \\
\hline
\raisebox{0pt}[12pt][6pt]{LEP2\cite{lep2}} &
\raisebox{0pt}[12pt][6pt]{$\bar e \gamma_\mu e \bar\ell \gamma^\mu \ell$} &
\raisebox{0pt}[12pt][6pt]{6.1} &
\raisebox{0pt}[12pt][6pt]{4.5} \\
\raisebox{0pt}[12pt][6pt]{} &
\raisebox{0pt}[12pt][6pt]{$\bar e \gamma_\mu \gamma_5 e \bar b 
\gamma^\mu \gamma_5 b$} &
\raisebox{0pt}[12pt][6pt]{4.3} &
\raisebox{0pt}[12pt][6pt]{3.2} \\
\hline
\raisebox{0pt}[12pt][6pt]{MFV\cite{mfv}} &
\raisebox{0pt}[12pt][6pt]{$\frac{1}{2}({\bar q}_L\lambda_u\lambda_u^\dagger
\gamma_\mu q)^2$} &
\raisebox{0pt}[12pt][6pt]{6.4} &
\raisebox{0pt}[12pt][6pt]{5.0} \\
\raisebox{0pt}[12pt][6pt]{} &
\raisebox{0pt}[12pt][6pt]{$H^\dagger {\bar d}_R \lambda_d\lambda_u
\lambda_u^\dagger \sigma_{\mu \nu}q_LF^{\mu\nu}$} &
\raisebox{0pt}[12pt][6pt]{9.3} &
\raisebox{0pt}[12pt][6pt]{12.4} \\
\hline
\end{tabular}
\end{center}
\end{table}

I will refer to the ``little'' hierarchy (LH)
problem as the tension between the
constraints on $\Lambda_{\rm LH}$ -- which imply that new-physics virtual effects
could only emerge at energies larger than 5--10 TeV -- and the no-fine-tuning
condition -- which requires the presence of new dynamics at the scale
$\Lambda_{\rm SM}$, below
the TeV. To some readers this could seem like a marginal problem, but I
believe it provides a very useful guideline in the search for the correct
theory beyond the SM. It is a typical LEP heritage, 
where any new-physics theory has to be confronted with very
precise data.

The ``little'' hierarchy between $\Lambda_{\rm SM}$ and $\Lambda_{\rm LH}$ 
is already giving us some lessons on the hypothetical theory beyond the SM.
First of all, new physics at $\Lambda_{\rm SM}$  is most likely weakly-interacting,
or else effective operators would be generated with $\Lambda_{\rm LH}\simeq
\Lambda_{\rm SM}$. Strong dynamics, if any, can only appear at scales larger than 
$\Lambda_{\rm LH}$. Moreover, even for weak dynamics, physics at $\Lambda_{\rm SM}$
should not induce (sizable) tree-level
contributions to effective operators. 
Recently there has been growing interest in constructing new theories that
describe the physics between $\Lambda_{\rm SM}$ and $\Lambda_{\rm LH}$, as I will
report in Secs.~\ref{sec4}, \ref{sec5}, and \ref{sec6}.

\section{Supersymmetry}\label{sec2}
Supersymmetry\cite{susy} provides an elegant solution to the ``big'' hierarchy
problem, by eliminating quadratic divergences with a symmetry principle.
In softly-broken supersymmetry, the superpartner masses (generated
by gauge-invariant terms) effectively play the r\^ole of $ \Lambda_{\rm SM}$.
Because of the absence of quadratic divergences, the validity of the 
theory can be extended to energy scales as large as $M_{\rm Pl}$, without
encountering naturalness problems. This extension has several welcome
consequences. It offers the possibility to 
link the SM to speculative
ideas about quantum gravity at $M_{\rm Pl}$. It provides a
viable setting for GUT's (with successful unification of gauge coupling
constants), a framework for neutrino masses, and for suppressing
proton decay. Finally, it allows an implementation of cosmological
mechanisms, such as inflation or 
baryogenesis, which benefit from extrapolations
of particle physics models to high energies.

Incidentally, 
the need for an extrapolation of the SM up to super-heavy scales has been
challenged by recent research aiming at constructing new models
whose validity cut-off is not very far from the electroweak scale. In
these scenarios, at first sight, the advantages of the energy ``desert''
are lost. Nevertheless, theoreticians have suggested new ways of recovering
some of the positive aspects. Just as an example, let me consider gauge
coupling unification. In theories with extra dimensions, Kaluza--Klein
excitations of SM particles can accelerate the running, possibly leading to a 
precocious gauge unification\rlap,\,\cite{ghe} although predictivity is lost in the
power running. Alternatively, one can consider an $N$-fold replication of the
SM gauge group, where the unification scale becomes 
$10^{13/N}$~TeV\rlap.\,\cite{stup}
Yet, one can abandon the usual tree-level expression for the weak mixing angle
in GUT's: $\sin^2\theta_W={\rm Tr}~I_3^2/{\rm Tr}~Q^2=3/8$ (where the trace
is over any GUT irreducible representation). Taking an electroweak group
$SU_3\times SU_2\times U_1$ broken to $SU_2\times U_1$, in the limit
${\tilde g}_2,{\tilde g}_1\gg {\tilde g}_3$ (where ${\tilde g}_i$ are the
coupling constants of the high energy group), one finds $\sin^2\theta_W=
1/4$\rlap.\,\cite{dim} This is much closer to the experimental value 
$\sin^2\theta_W =0.231$ than $3/8$, and therefore little running is needed.

Aside from this connection to super-high energies, 
supersymmetry has 
several interesting and successful features in the low-energy domain.
{\it (i)} Gauge-coupling unification leads to a prediction of a SM parameter:
$\alpha_s(M_Z)=0.124$ (for typical supersymmetric thresholds at 1 TeV)
and 0.130 (for thresholds at 250 GeV). Given the intrinsic uncertainty
of GUT thresholds, this
is consistent with the PDG value $\alpha_s(M_Z)=0.117\pm 0.002$\rlap.\,\cite{pdg}
{\it (ii)} Electroweak symmetry is triggered radiatively by the top Yukawa
interaction,
with the broken group dynamically chosen (color $SU_3$ is preserved
since squarks do not develop negative square masses, while weak $SU_2$
is broken).
{\it (iii)} The Higgs boson is predicted to be lighter than about 130 GeV,
compatibly with the indications from electroweak precision data.
{\it (iv)} The ``little'' hierarchy is satisfied, if one assumes $R$-parity
conservation. In this case, supersymmetric virtual effects are loop-suppressed
and $\Lambda_{\rm LH}\sim 4\pi \Lambda_{\rm SM}$. 
{\it (v)} There is a natural candidate for cold dark matter, if
$R$-parity is conserved.

\begin{figure}
\center
\psfig{figure=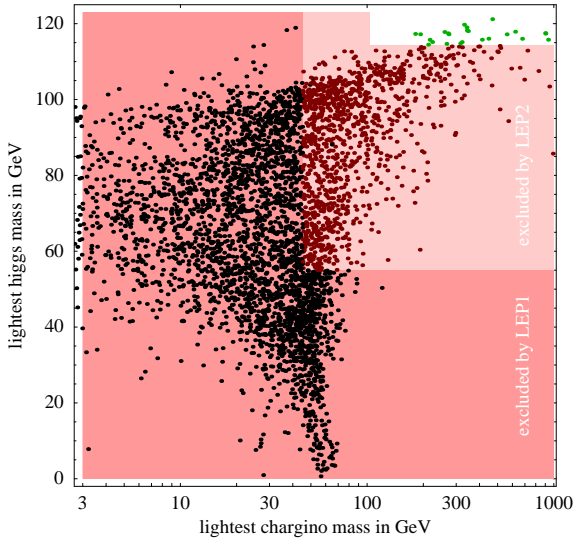,width=8.0truecm}
\caption[]{Scatter plot of the chargino and Higgs mass obtained by
sampling the parameter space of minimal supergravity. The regions
excluded by LEP1 and LEP2 searches are shown. This figure was
produced by A.~Strumia, updating earlier results\rlap.\,\cite{sd}}
\label{fig1}
\end{figure}

These positive aspects of low-energy supersymmetry are well known, but let
me spend a few words on the negative aspects. The first
obvious drawback of low-energy
supersymmetry is that no superpartner has been observed. This is in 
contradiction with the requirement of less than 10\% fine-tuning, which 
predicted a discovery at LEP2\rlap.\,\cite{fine} In conventional models of
low-energy supersymmetry, the most stringent direct bounds come from
chargino ($m_{\chi^+}>103.5$~GeV)\rlap,\,\cite{chi} charged slepton ($m_{\tilde 
e_R}>99.4$~GeV)\rlap,\,\cite{slep} and
Higgs searches. The SM Higgs mass bound of 114.4~GeV\cite{higgs} 
applies only in the limit
of a heavy pseudoscalar, while the general bound is 91.0~GeV\cite{shiggs} 
(the final 
combined LEP analysis on the supersymmetric
Higgs has not appeared yet).   
These limits imply a fine-tuning of the order of the per cent. This
is illustrated in Fig.~\ref{fig1}, which shows the values of Higgs and
chargino masses obtained by a random sample of supersymmetric
models with universal boundary conditions at the GUT scale and correct
electroweak breaking\rlap.\,\cite{sd} Less than 1\% of the points have simultaneously
$m_{\chi^+}> 103$~GeV and $m_H>114$~GeV (for pseudoscalar Higgs masses
close to $m_H$, the bound becomes weaker but the allowed region of
Fig.~\ref{fig1} remains essentially the same).

This is a very disappointing result of supersymmetry, leading to three
possible explanations. {\it (i)} Supersymmetry is not realized at the
weak scale. {\it (ii)} Low-energy supersymmetry is correct, but parameters
are fine-tuned at the level of the per cent (and LHC will discover
the superpartners). In this case, you may be relieved to know that
there is one example in nature of an
accidental correlation of this order of magnitude. The apparent sizes
of the Sun and the Moon in the sky are roughly equal. If you
applied the naturalness
criterion to conclude that their actual sizes are roughly equal too,
you would be quite wrong. Their radius and distance to the Earth are
``fine-tuned'' to give, on average, $(\theta_{\rm Sun}-\theta_{\rm
Moon})/\theta_{\rm Sun}\sim 3\%$, where $\theta_{\rm Sun,Moon}$ are
the solar and lunar angular sizes.
{\it (ii)} Low-energy supersymmetry exists, but its realization is
different from what we had imagined. Let me expand on this. 

The origin of the fine-tuning is linked to the success of supersymmetry
to trigger the electroweak breaking. Indeed, from the Higgs mechanism we know
that $M_Z^2=-g^2 \mu_H^2/\lambda$, where $\mu_H^2$ and $\lambda$ are
the coefficients of the quadratic and quartic terms in the Higgs 
potential. In supersymmetry, $\lambda \sim g^2$, and therefore
$M_Z^2\sim- \mu_H^2$.
The effective $\mu_H^2$ has
a tree-level direct contribution from the soft terms and a one-loop
negative contribution roughly proportional to the stop mass (which, in 
turn, is usually determined by the contribution proportional to the
gluino mass). In the conventional supergravity approach, the 
one-loop contribution
has a large logarithm and it overcomes the tree-level effect. Therefore
$|\mu_H |$ is of the order of the supersymmetric masses, which are expected
to lie close to $M_Z$. One may think that the situation improves in
gauge-mediated models\rlap,\,\cite{gmed} since the logarithm can be made small with
sufficiently light
messengers. However, this is not the case,
because gauge mediation gives a boundary condition for the stop mass
at the messenger scale, which is significantly larger than the tree-level
contribution to $\mu_H^2$. Actually, in gauge mediation, the fine-tuning
is usually more acute, because the right-handed selectron is a factor
of about 2 lighter than the Higgs mass parameter, and because of the
limit on the visible decay of the lightest neutralino\rlap.\,\cite{sd} 
The situation could
improve in models where the stop and gluino masses are not large (with 
respect to the other soft masses) or in models with a low supersymmetry-breaking
scale\rlap,\,\cite{espp} or in models where
$\mu_H^2$ is truly a one-loop factor smaller than all supersymmetric masses
(see for instance ref.\cite{loop}).

The second drawback of low-energy supersymmetry is the lack of predictivity
in the supersymmetry-breaking sector, which translates into the 
``supersymmetric flavor
problem'' (i.e. the large flavor violations present with generic
soft terms). We are now aware of various schemes, alternative to the 
conventional supergravity scenario, which are more predictive and
address the flavor problem: gauge mediation\rlap,\,\cite{gmed} 
anomaly mediation\rlap,\,\cite{ano} gaugino mediation\rlap.\,\cite{gau} 
However, at present,
we cannot say that one scheme is preferable to the others. For instance,
in operator-based schemes (like supergravity) the Higgs mixing mass $\mu$
can be properly accounted for\rlap,\,\cite{giu} while in other schemes where
the soft terms are calculable, the Higgs mass parameters $\mu$ and $B\mu$
typically appear at an undesired order in perturbation theory. Luckily,
the different schemes usually have quite distinct features both
in the mass spectrum and in the nature of the lightest supersymmetric
particles, leading to distinct signals at colliders. Therefore, in the case
of a discovery, the question of the correct scheme of supersymmetry-breaking
terms can be settled experimentally.

Recently many new models have been proposed, where supersymmetry at the
weak scale is implemented in unconventional ways. These proposals make use of
new ingredients, coming from extra dimensions. Before discussing them, let me
first introduce extra dimensions. 

\section{Extra Dimensions}\label{sec3}
The hierarchy problem can motivate the existence of experimentally-accessible
extra dimensions\rlap.\,\cite{add,rs1} The scenarios are by now well known: our 
three-dimensional space is embedded in a larger D-dimensional space-time.
While SM fields are confined on the three-dimensional brane, gravity
is described by the geometry of the full space. The crucial assumption is that
the fundamental Planck scale of the $D$-dimensional 
theory $M_D$ is of the order
of the weak scale, avoiding any hierarchy. Any distance scale shorter
than TeV$^{-1}$ (associated with Newton's constant, with GUT's, right-handed
neutrinos, etc.) has to be explained by some geometrical property.

For flat and compact extra dimensions, the observed weakness of gravity is
translated into a large value of the compactification radius $R$, since
the ordinary Planck mass is given by\cite{add} 
$M_{\rm Pl}\sim R^{\delta/2}M_D^{1+
\delta/2}$, where $D=4+\delta$. The interpretation is that we observe a weak
gravitational force not because gravity is intrinsically weak ($M_D\sim$ TeV),
but because of the small overlap between our world and the
graviton wave function, which is spread in a very large volume.

For spaces with non-factorizable metrics, the large hierarchy $M_{\rm Pl}/M_W$
can be reproduced with much smaller values of $R$, exploiting the strong
functional dependence in the warp factor. In this case, the interpretation
is that, because of the non-trivial geometry, the zero-mode graviton 
wave function is peaked on a brane far from ours and only its
exponential tail overlaps with our world, leading to the observed weakness
of gravity. The hierarchy is the result of the familiar gravitational 
redshift in general relativity. A photon that climbs out of a gravitational
potential sees its measured energy reduced. 
It is known that for time-independent
metrics with $g_{0j}=0$, the product $E\sqrt{|g_{00}|}$ is conserved, rather
than the energy $E$. Therefore, a photon emitted at a distance $r$ from
a spherically symmetric gravitational source is observed at infinity
to be redshifted by an amount $(E_{obs}-E_{em})/E_{em}=\sqrt{|g_{00}|}
-1\simeq -G_N M/r$, since the Schwarzschild metrics generated by a mass $M$
has $g_{00}=1-2G_N M/r$. Similarly, in the Randall--Sundrum set-up, masses
in the hidden (Planck) brane are blueshifted by a warp factor, when measured on
the observable (TeV) brane, as a result of the non-trivial gravitational field.

Although the scenarios with a quantum-gravity scale at the TeV have various
theoretical and cosmological difficulties, they represent a wonderful
possibility for new-physics searches at future colliders. At the LHC, graviton
emission is observed in jet plus missing-energy events\cite{noi,alt} 
(for flat dimensions)
or as resonances in Drell--Yan\cite{tom} (for warped dimensions). Tree-level
graviton exchange leads to an effective dimension-8 operator $T_{\mu \nu}
T^{\mu \nu}$, which predicts correlated signals in diphoton and dilepton
final states\rlap.\,\cite{noi,var} Graviton loops give rise to dimension-6 operators,
with a flavor-universal axial-vector contact interaction playing a special 
r\^ole\rlap.\,\cite{str} Because of the 
lower dimensionality of the induced operators,
loop effects are generally more important than tree-level effects, unless
the short-distance behavior of gravity is modified at an energy scale
significantly smaller than $M_D$. The radion, a scalar field contained
in the extra-dimensional graviton, can mix with the Higgs boson,
giving rise to interesting variations in the Higgs searches\rlap.\,\cite{noi2}

The theoretical description of the experimental signals described above
relies upon linearized gravity, whose validity ends as the relevant
energy of the process ($\sqrt{s}$) approaches the fundamental gravity
scale $M_D$, and we enter the region where experiments are directly sensitive
to the underlying quantum-gravity theory. However, it is interesting that 
there exists another kinematical regime where the theoretical description
is tractable. This is the transplanckian region $\sqrt{s}\gg M_D$, 
where a semi-classical
description is appropriate. This can be understood by noticing that,
in the transplanckian region, the
classical length associated with a gravitational scattering --
the Schwarzschild radius $R_S\sim (G_D\sqrt{s})^{1/(\delta+1)}$, where
$G_D\sim M_D^{-(\delta +2)}$ is the $D$-dimensional Newton constant
-- becomes larger than the Planck length $M_D^{-1}$ associated to 
quantum-gravity effects. 

\begin{figure}
\center
\psfig{figure=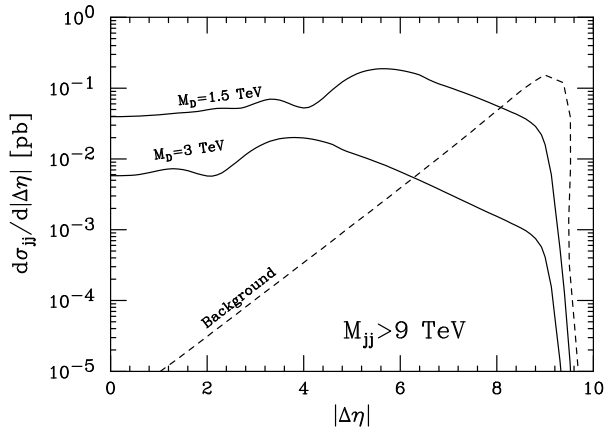,width=8.0truecm}
\caption{The di-jet differential cross section 
$d\sigma_{jj}/d|\Delta\eta |$ versus the jet rapidity separation
$\Delta\eta $ at the LHC
from transplanckian
gravitational scattering for $\delta=6$, $M_D=1.5$~TeV and $3$~TeV.
Cuts are chosen such that the  
di-jet invariant mass $M_{jj}>9$~TeV, and both jets have $|\eta | <5$ 
and $p_T>100$~GeV.
   The
dashed line is the expected background from QCD.}
\label{fig2}
\end{figure}

When the collision impact parameter $b$ is
larger than $R_S$, the non-linearity of the gravitational field is
negligible, and the gravitational scattering, although non-perturbative,
can be theoretically calculated. This leads to a prediction at hadron colliders
for two-jet
final states at large invariant mass and large rapidity separation
with characteristic distributions\cite{noi3} (an example is shown in 
Fig.~\ref{fig2}). 
The peculiarity in the rapidity
distribution comes from a diffractive pattern associated with a new scale 
in the gravitational potential $b_c\sim (G_D s)^{1/\delta}$, which exists
in any dimensions different from 4 (i.e. $\delta \ne 0$) -- the familiar
Coulomb-like potential is actually very special. These features can be used
to confirm the gravitational nature of possible discoveries of new physics.

As the impact parameter $b$ approaches the Schwarzschild radius $R_S$,
the gravitational field becomes strong, and analytic calculations cannot
be done. However, there are solid indications\cite{sod} that, when 
$b<R_S$, black holes are formed\rlap.\,\cite{bh} This opens the exciting possibility
that (unstable) black holes can be observed at future colliders, with
spectacular experimental signals.

I should remark that given the present bounds on $M_D$ from direct searches,
and the ``limited'' energy available at the LHC, the window for observing
transplanckian phenomena is rather narrow and, more importantly, the
separation between classical and quantum scale may not be sufficient to
avoid pollution from quantum-gravity effects. Although it is of course
important to pursue these searches at the LHC, transplanckian phenomena
can offer a good motivation to consider future hadron colliders, such as
the VLHC, operating at $\sqrt{s}\sim $ 100--200 TeV.

As we have considered extra-dimensional gravity, it is possible that also
SM particles live in (part of) the extra-dimensional compactified 
space\rlap.\,\cite{anto} Collider experiments can look for their 
Kaluza--Klein (KK) excitations. 
If only gauge bosons live in five dimensions, while fermions are conventional,
a combination
of present direct and indirect searches yield a 95\% CL lower limit 
on the compactification scale $M_C$ (the mass of the first excited mode)
of 6.8~TeV\rlap.\,\cite{chu} Searches at the LHC have a sensitivity reach of up to 
13--15 TeV. Much weaker bounds ($M_C>0.3$~TeV) apply to the case
of universal extra dimensions\rlap,\,\cite{uni} where all SM fields live in five 
dimensions. This is because, after compactification, the momentum 
conservation along the fifth dimension corresponds to a conserved
KK number. This plays a  r\^ole analogous to $R$-parity conservation 
in supersymmetry. Indeed KK particles can only be pair-produced and
cannot contribute to SM processes by virtual tree-level exchange. Moreover,
the lightest KK mode, corresponding to the hypercharge gauge boson, 
is stable and is a possible candidate for cold dark matter\rlap.\,\cite{cdm} 

\section{Supersymmetry Breaking and \\ Extra Dimensions}\label{sec4}
At the end of Sec.~\ref{sec2}, I anticipated that extra dimensions 
bring new elements in the construction of low-energy supersymmetric models.
Let me briefly discuss these new ingredients, which are related to the issues 
of supersymmetry and 
gauge-symmetry breaking, orbifold projection, and the AdS/CFT correspondence. 

Consider a field $\Phi$, which lives in a five-dimensional space with the
fifth dimension $y$ 
compactified on a circle of radius $R$. Because of periodicity,
the field can be expanded in discrete Fourier modes in $y$, which correspond
to the familiar KK tower of four-dimensional 
fields with masses $m_n=n/R$, for integers $n$.
Let us now impose the non-trivial boundary condition
that the field $\Phi$ picks up
a phase as it goes around the circle:
\beq
\Phi (x,y+2\pi R)=e^{2\pi i Q_\Phi} \Phi (x,y).
\label{kkk}
\eeq
The KK expansion consistent with condition (\ref{kkk}) is
\beq
\Phi(x,y)=e^{iQ_\Phi y/R}\sum_{n=-\infty}^{+\infty} e^{iny/R}\Phi_n (x).
\label{decc}
\eeq
The mass of the $n$-th mode (i.e. its momentum along the fifth 
coordinate) is shifted by the boundary condition to the value
$m_n=(n+Q_\Phi)/R$. Since the mass spectrum
of different fields can be shifted in different ways, symmetries of the
five-dimensional theory can be hidden if we look at its four-dimensional 
truncated version, obtained by keeping only light modes. If the
phase $Q_\Phi$ corresponds to an $R$-charge (i.e. if particles inside
the same supermultiplet have different boundary conditions), then supersymmetry
is broken in the four-dimensional effective low-energy theory. This is
the Scherk--Schwarz 
symmetry-breaking mechanism through boundary conditions in the extra 
dimensions\rlap.\,\cite{ss} Notice that the mechanism is intrinsically non-local,
since it involves the global structure of the compactified space. Therefore,
if we look at small distances (smaller than $R$), any symmetry-breaking
effect should disappear. This is a very useful property. For instance,
in order to stabilize the weak scale it is essential 
to maintain the cancellation of
quadratic divergences in the ultraviolet, 
and therefore we want to recover supersymmetry at short distances.

The compactification of the fifth dimension into a circle was obtained
by identifying the point $y$ with $y+2\pi R$. We can further restrict
the space by identifying $y$ with $-y$: the circle collapses into a segment.
The KK decomposition (take Eq.~(\ref{decc}) with $Q_\Phi =0$ for simplicity)
can be written as $\Phi=\Phi^{(+)}+i\Phi^{(-)}$, where
\bea
\Phi^{(+)}(x,y) &=& \sum_{n=0}^{\infty}\left[ \Phi_n(x)+\Phi_{-n}(x)
\right] \cos (ny/R) \cr
 \Phi^{(-)}(x,y) &=& \sum_{n=1}^{\infty}\left[ \Phi_n(x)-\Phi_{-n}(x)
\right] \sin (ny/R).
\eea
The fields $\Phi^{(+)}$ and $\Phi^{(-)}$ are even and odd, respectively,
under the symmetry $y\to -y$. However, only the even tower has a massless
zero mode ($n=0$). This projection into an orbifold (essentially
a manifold with 
boundaries) is very useful in model building. In particular\rlap,\,\cite{orb} 
starting from
vector-like fermion representations in extra dimensions, one 
can retain only some chiral
components in the truncated four-dimensional theory, where all massive modes
have been integrated out, and construct realistic models for weak 
interactions. Combinations of Scherk--Schwarz boundary conditions and
orbifold projections give rise to many interesting possibilities for
reducing the symmetry and the particle content of extra-dimensional
theories.

The next ingredient I want to discuss is the AdS/CFT correspondence\rlap,\,\cite{ads}
which is the conjecture that properties of conformal field theories
in $D$ dimensions are related to those of $(D+1)$-dimensional theories
with gravity in anti-de Sitter space. It has also been suggested that
the non-compact version of the RS model (RS2)\cite{rs2} is 
dual to a four-dimensional
strongly-coupled conformal theory with gravity\rlap.\,\cite{conf}
One can take advantage 
of these speculations to extract
some physical information on the structure of the RS1 set-up with two branes
in a slice of AdS$_5$, discussed in Sec.~\ref{sec3}\rlap.\,\cite{cons} 

The line element of the RS set-up is given by
\beq
ds^2=\exp\left( {-\frac{2y}{L}}\right) dx^\mu dx_\mu +dy^2,
\label{line}
\eeq
where $y$ is the extra coordinate and $L$ is the AdS radius.
From Eq.~(\ref{line}) we observe that a scale
transformation $x\to \xi
x$ can be reabsorbed by a shift of the coordinate $y$. This is suggestive
of the holographic interpretation of the fifth coordinate as the
renormalization scale of the four-dimensional theory. 
The hidden (Planck) brane 
corresponds to the ultraviolet cut-off of the conformal theory and
to the addition of gravity.
Indeed, gravity decouples from the four-dimensional theory as the Planck
brane is moved to infinity. The visible (TeV) brane is the infrared cut-off,
where the observable fields live. It corresponds to a spontaneous breaking 
of the conformal symmetry. Local gauge symmetries in the 
bulk of the five-dimensional
theory are matched to global symmetries of the conformal theory.
This connection between higher-dimensional and quasi-conformal theories
could be very useful for gaining new insights on some properties of the 
strongly-coupled regime, which can be accessed by perturbative calculations
in the dual theory.  

The ingredients presented above have been used by many authors to construct
new implementations of supersymmetry at the weak scale. 
Although I find many of these attempts quite interesting, I do not intend 
to give a review of these proposals here. Instead I will take only two 
examples, to show the use of the higher-dimensional ingredients.

The first example\cite{bar} is based on the extension of the supersymmetric
model into five dimensions, one of which is compactified on a circle 
with two orbifold projections, $S^1/(Z_2\times Z_2)$. In five dimensions, there
are two supersymmetries. Each boundary of the orbifold breaks one supersymmetry
and therefore the effective low-energy theory is non-supersymmetric.
However, supersymmetry is recovered at distances smaller than the 
compactification radius $R$, leaving only a {\it finite} contribution
to the Higgs mass, and predicting\cite{bar} $m_H=127\pm 10$
GeV. It is remarkable that one is able to compute a SM parameter 
(most new-physics theories add new parameters instead of
predicting the values of SM quantities!). Unfortunately a quadratic sensitivity
to the ultraviolet cut-off is actually
introduced by a one-loop contribution to the
Fayet--Iliopoulos term\rlap.\,\cite{nill} 
Also the contribution to the $\rho$ parameter
is too large, unless we tune the coefficient of an ultraviolet-sensitive
operator. Finally unitarity is violated at a scale of 1.7 TeV, where an
ultraviolet completion is needed. Variations of this model have been 
proposed\rlap,\,\cite{vbar} in which some of these problems can be alleviated, in
particular by raising the cut-off into the 5--10 TeV region.

In this model we can clearly appreciate the attractive features of
Scherk--Schwarz supersymmetry breaking and orbifolding. Note also that,
from the point of view of collider searches, these models are completely
distinct from the conventional low-energy supersymmetric ones. Indeed, here
there is only one Higgs doublet and two superpartners for each SM particle
(the remnant of supersymmetry in five dimensions). The lightest supersymmetric
particle
is a stable (or metastable, if small $R$-parity violations are included)
stop with a mass of about 210 GeV.

The second example\cite{pom} I want to present 
exploits the possibility of coexistence 
of supersymmetric and non-supersymmetric sectors in extra 
dimensions\rlap.\,\cite{lut}
Let us consider a supersymmetric RS1 set-up where the Higgs sector lies on
the TeV brane and the other SM multiplets live in the bulk. A source
of supersymmetry breaking is localized on the Planck brane. The bulk fields
are directly coupled to the supersymmetry-breaking sector and superpartners
(squarks, sleptons, and gauginos) 
acquire masses of the order of $M_{\rm Pl}$: supersymmetry is badly broken.
However the Higgs sector can only feel supersymmetry-breaking effects 
through loops involving brane-to-brane mediation of bulk particles. Since
this corresponds to a non-local interaction, the contribution is finite.
Moreover, because of the gravitational redshift discussed in Sec.~\ref{sec3},
the Higgs sector feels that the supersymmetry-breaking source is a warp factor
smaller than $M_{\rm Pl}$. Using the AdS/CFT correspondence, we can interpret
the theory in four dimensions. The SM fields are elementary, while the Higgs
sector corresponds to bound states of a quasi-conformal field theory.
The electroweak scale is generated at the one-loop level, and therefore
$\langle H \rangle \sim (1/4\pi) L^{-1}$, where $L$ (assumed to be TeV$^{-1}$) 
is the size of the 
bound states. The loop factor is very welcome to explain the little
hierarchy discussed in Sec.~\ref{sec1}. The collider phenomenology is
again quite distinct from usual low-energy supersymmetry. Only the
Higgs has supersymmetric partners (higgsinos), while a tower of CFT
bound states will appear at energies of the order of $L^{-1}$. 
This interpretation allows a connection of this set-up with models of
composite Higgs bosons\rlap.\,\cite{dpom} 
Unfortunately,
precision electroweak data set a lower bound on $L^{-1}$ of the order of
9 TeV, implying a certain degree of fine-tuning in the generation of the
weak scale. I hope this example has clarified how AdS/CFT can be used in
model building to extract information in theories with strong interaction
and composite Higgs bosons.

In conclusion, extra dimensions have brought in the game new ingredients
for low-energy supersymmetric models. Supersymmetry could reveal itself
at colliders in a variety of ways: not only with missing energy, as
in supergravity models, but also accompanied by leptons, or photons,
or stable charged particles (as in gauge mediation\cite{gmed}) or
with nearly-degenerate $\tilde W^{\pm,0}$ (as in anomaly 
mediation\cite{ano2}),
or with new characteristic signals (stable stop, partially supersymmetric
spectrum, etc.).
 
\section{Electroweak Breaking and \\ Extra Dimensions}\label{sec5}

The ingredients that I have discussed in the previous section can also be 
used to break gauge symmetries, and it is therefore a logical possibility 
to use extra dimensions to break the electroweak group.

The hierarchy problem is solved by a symmetry that forbids the Higgs mass
term. Supersymmetry (in conjunction with chiral symmetry) is an example.
Another example could be gauge symmetry $A_\mu \to A_\mu +\partial_\mu \Lambda$
that forbids a mass
term $m^2A_\mu^2$, but the Higgs boson is not a gauge particle. However, 
this is
true in four dimensions, but not necessarily in more dimensions. A gauge boson
in $D=4+\delta$ dimensions can be 
decomposed as $A_M=(A_\mu , A_j)$, $j=1,\dots ,\delta$, where $A_j$ are scalar
fields under the Lorentz group. The extra-dimensional components could
be interpreted as Higgs bosons\rlap.\,\cite{hdim,hoso} 
This provides a Higgs--gauge 
unification completely analogous to the old version of the KK
unification of electromagnetism and gravity, in which photon and graviton are
different components of the same field -- the five-dimensional metrics.

Since, unlike the gauge boson, the Higgs 
does not transform under the gauge group as the adjoint
representation, one has to rely on the orbifolding discussed in
Sec.~\ref{sec4} to project out unwanted states. More difficult is
to generate the Higgs Yukawa and quartic couplings without reintroducing
quadratic divergences. This problem is typically harder than the similiar
one encountered in Little Higgs theories (discussed in Sec.~\ref{sec6}),
since the structure of the Higgs interactions is tightly constrained by
the embedding in of extra dimensions. Nevertheless, there has been 
progress towards the construction of realistic models\rlap.\,\cite{pro}

Following a different approach, one can try to break the electroweak group
by appropriate boundary conditions of the gauge bosons in the extra dimensions,
and eliminate the Higgs altogether. Of course, in this case, one expects
KK excitations for the $W$ and $Z$ bosons with masses too low to satisfy
present collider bounds\rlap.\,\cite{mur1} By warping the extra dimension, one
can lift the masses of the excited KK modes with respect to the zero modes
(to be identified with ordinary gauge bosons) and make these particles
phenomenologically
acceptable\rlap,\,\cite{ssun,mur2} opening a path for the construction of realistic models
for electroweak breaking with no Higgs (neither fundamental nor composite).
Notice that these proposals differ substantially
from the case of a Higgsless SM where
the gauge symmetry is non-linearly realized. In the latter case unitarity
in scattering processes involving longitudinally-polarized $W$ bosons
is violated at an energy scale $G_F^{-1/2}$, below the TeV. On the contrary,
violation of unitarity in higher-dimensional theories is postponed to
energy scales in the 10~TeV region\rlap.\,\cite{duan} This is sufficient to allow
these theories to pass the test of the little hierarchy (see 
Sec.~\ref{sec1}). Effectively, the tower of gauge boson KK modes (partly)
plays the r\^ole of the Higgs boson in modifying the cross section high-energy
behavior.

At present the main obstacle for Higgsless extra-dimensional theories
seems to come from electroweak precision data. Contributions to the 
$\rho$ parameter can be made small by imposing a gauge custodial symmetry
in the bulk (which has the holographic interpretation of a global
custodial symmetry in the four-dimensional theory, see Sec.~\ref{sec4}).
However, contributions to the oblique parameter $S$\cite{pesk} (or 
$\epsilon_3$\cite{balt}) appear to be positive and unacceptably 
large\rlap.\,\cite{ratt}
  
\section{Little Higgs}\label{sec6}
We have seen that supersymmetry and gauge symmetry are two examples 
of principles that can be used to forbid a mass term for the Higgs
boson and solve the hierarchy model. A third example is a non-linearly
realized global symmetry, where the Higgs has to be identified with
a Goldstone boson. 

Consider a scalar field $\Phi$, transforming under an abelian global
symmetry as $\Phi \to e^{ia}\Phi$, and assume that the symmetry is 
spontaneously broken by $\langle \Phi \rangle =f$. I can parametrize
the complex field $\Phi$ in terms of two real fields $\rho$ and $\theta$
as $\Phi=(\rho +f)e^{i\theta /f}/\sqrt{2}$. The symmetry transformation 
properties are $\rho \to \rho$ and $\theta \to \theta +a$. If I identify
$\theta$ (the Goldstone boson) with the Higgs, the symmetry forbids
a mass term $m^2 \theta^2$.

The non-trivial part of the problem is to generate the gauge, Yukawa,
and self-interaction of the Higgs (which are non-derivative couplings
and therefore forbidden by the symmetry) without reintroducing quadratic
divergences. Some (not entirely successful) attempts to construct 
realistic models for the Higgs as a pseudo-Goldstone boson were made in the 
past\rlap.\,\cite{pseud} 

Recently, a much less ambitious program was proposed, which is known
under the name of ``Little Higgs''\rlap.\,\cite{litt} The proposal is to solve only
the little hierarchy problem; in other words, one searches for a description
valid only up to 10 TeV or so. Recall, from Sec.~\ref{sec1}, that the
big hierarchy problem is formulated in terms of one-loop corrections to the 
Higgs mass:
\beq
\delta m_H^2\sim \frac{G_F}{\pi^2} m_{\rm SM}^2 \Lambda_{\rm SM}^2.
\eeq
If we require this correction not to exceed the masses of the
SM particles $m_{\rm SM}$, we obtain
\beq
 \Lambda_{\rm SM}<\frac{\pi}{\sqrt{G_F}}\sim {\rm TeV}.
\eeq
Suppose that at the scale $\Lambda_{\rm SM}$ $(<{\rm TeV})$ new physics
cancels the one-loop power divergences. Then we are left with one-loop
logarithmic divergences and two-loop power divergences of the form
\beq
\delta m_H^2\sim \frac{G_F^2}{\pi^4} m_{\rm SM}^4 \Lambda^2.
\eeq
The same naturalness argument now implies
\beq
\Lambda < \frac{\pi^2}{G_Fm_{\rm SM}}\sim 10~{\rm TeV},
\eeq
which is an energy scale of the order of $\Lambda_{\rm LH}$ (defined in
Sec.~\ref{sec1}). Therefore, cancelling only one-loop divergences is
just sufficient to postpone the big hierarchy problem beyond the scale
of the little hierarchy problem, thus being consistent with the 
phenomenological constraints.

In order to achieve the cancellation, at least at one-loop order, we can
use the mechanism of ``collective breaking''. There are two (or more)
approximate global symmetries under which the Higgs is a Goldstone boson.
Therefore, each symmetry protects the Higgs mass and one requires that
no single term in 
the Lagrangian simultaneously break all these symmetries. More than one term
is necessary to generate the Higgs mass, and therefore the corresponding 
Feynman diagram has more than one loop.

The collective breaking can be achieved with gauge-group replication and,
in this sense, the Little Higgs can be interpreted as a model with discrete
extra dimensions through deconstruction\rlap.\,\cite{deco} The model-building recipe
is the following. One starts from a non-linear sigma model of the 
Goldstone bosons in the coset $G/H$. The group $G$ has a weakly-gauged
subgroup  $G_1\times \dots \times G_n$, where $n\ge 2$. Each of the gauge 
groups $G_j$ preserves a different non-linear global symmetry under which
the Higgs transforms as a Goldstone boson. The SM group is a subgroup
of $G_1\times \dots \times G_n$ which breaks all the global symmetries.
Divergent contributions to the Higgs mass necessarily involve the
gauge couplings from all the $G_j$ groups and are therefore generated only
at the $n$-th loop:
\beq
\delta m_H^2\sim \frac{g_1^2}{(4\pi)^2}\dots \frac{g_n^2}{(4\pi)^2}\Lambda^2.
\eeq

Instead of using gauge-group replication, one can obtain collective breaking
by replicating the field content. Consider the example\cite{kapp} of an
$SU(N)$ gauge group with two scalar fields $\Phi_{1,2}$
in the fundamental representation. Take a scalar potential of the form
$V(\Phi^\dagger_1\Phi_1,\Phi^\dagger_2\Phi_2)$ such that both $\Phi_{1,2}$ 
acquire vacuum expectation values, spontaneously breaking the
$SU(N)$ gauge symmetry. In the limit in which you turn off the gauge coupling
to $\Phi_1$, the theory has a local $SU(N)$ under which only $\Phi_2$
transforms, and a global $SU(N)$ under which only $\Phi_1$
transforms. Both symmetries are spontaneously broken and the spectrum
contains a Goldstone
boson. In the limit in which you turn off the gauge coupling
to $\Phi_2$, the situation is analogous after replacing the indices 1 and 2,
and therefore we still find a massless Goldstone boson. This means that
power-divergent contributions can only appear in diagrams with gauge
couplings to {\it both} $\Phi_1$ and $\Phi_2$, i.e. at two loops:
\beq
\delta m_H^2\sim \frac{g^4}{(4\pi)^4}\Lambda^2.
\eeq 

Along these lines, for both  gauge-group and field replications, many models
have been constructed. They are quite interesting although, admittedly,
rather elaborate\rlap.\,\cite{minh,kapp,mdls} Effectively, the common 
operative mechanism that leads
to the one-loop cancellation of power divergences is the introduction,
at the scale $f\sim \Lambda_{\rm SM}$, of new degrees of freedom. One-loop
diagrams from each SM particle are compensated by one-loop
diagrams involving a new particle, much in the same fashion as supersymmetry
cancels quadratic divergences. The peculiarity, however, is that 
in Little-Higgs theories the cancellation occurs between particles with
the same spin. For instance the top-quark contribution to the Higgs mass
is cancelled by a loop of a new vector-like charge-2/3 fermion with 
a dimension-five 
effective coupling with the Higgs. The $W$ and $Z$ contribution
is cancelled by new electroweak triplet and singlet gauge bosons with
negative couplings to the Higgs. The content of scalar particles is more 
model-dependent and, in some cases, new electroweak triplet scalar fields
are necessary for the cancellation. All these new degrees of freedom
represent the signature of Little-Higgs theories, which can be searched 
for in future colliders, in particular the LHC\rlap.\,\cite{phlh}  

Tevatron searches for new gauge bosons and LEP precision data (in particular
$\Delta \rho$ contributions from the new top-like fermion, from 
mixing with the new gauge bosons and, possibly, from the scalar triplet) 
provide significant constraints on
Little-Higgs models. In the minimal model of Arkani-Hamed {\em et al.}\rlap,\,\cite{minh} the
symmetry-breaking scale $f$ can be, at best, as low as 5~TeV\rlap.\,\cite{limt}
This leads to a lower bound on the mass of the new top-like quark
$m_{t^\prime} >2\sqrt{2}(m_t/v)f=$ $14~{\rm TeV}(f/5~{\rm TeV})$.
Comparing with the top contribution to the Higgs mass, we find that this
implies a fine-tuning of at least one part in a thousand. Variations
of the minimal model can significantly reduce the amount of 
fine-tuning needed. 

The validity of the Little-Higgs theories extends up to a region of about
10~TeV, where we expect an embedding into a more fundamental region.
The construction of such an ultraviolet completion is still an open and 
compelling question\rlap.\,\cite{nels} 

\section{Conclusions}\label{sec7}

During the last few years we have significantly enriched our basket of
theories for the electroweak scale, substantially deepened our understanding
of them, and constructed many previously-undiscovered variations.
Nevertheless, none of the known models
is fully satisfying or totally free from fine-tuning. The excessive
fine-tuning may well be
just a fortuitous accident, or it is a sign that some theoretical ingredient
is still missing (or, possibly, that we are on a completely wrong track!)

A great distinction exists between theories with and without a ``desert''.
I define ``desert'' the case 
in which some new dynamics modify the ultraviolet behavior of the Higgs
mass parameter below the TeV, but then physics can be extrapolated up
to a very large energy scale without inconsistencies. Conventional low-energy
supersymmetry is the best known example.
As previously discussed,
the desert scenario has some indisputable advantages.
\begin{itemize}
\item It allows a connection between SM physics and quantum gravity
or other speculative
theories at very short distances, such as GUT's or strings.
\item It offers the possibility of predicting some SM parameters
($\alpha_S$, $m_b/m_\tau$) through GUT relations. In particular, 
gauge-coupling unification in low-energy supersymmetry is certainly the most
remarkable information on new physics we have.
\item It naturally embeds the existence of a new mass
scale $\Lambda$, as suggested by the evidence for neutrino masses and for
a new dimension-5 operator $(1/\Lambda)\ell
\ell HH$ to be included in the SM Lagrangian.
\item It allows a natural suppression of fast proton decay and
unwanted flavor violations.
\item It provides a set-up for interesting cosmological theories
(inflation, baryogenesis, etc.), which require an extrapolation of particle
physics to very small distances. 
\end{itemize}

However, for each of these points, 
alternative (and more or less attractive)
explanations have been proposed in scenarios with a low cut-off.
In my presentation I did not have the time to enter into the various solutions,
but at present it is fair to conclude that the verdict for desert versus
non-desert is still pending.

The interest in constructing theories that extend the SM at the TeV, but
are valid only up to the 
10 TeV region, has both experimental and theoretical roots.
The experimental
LEP data have convinced us that the scale at which new-physics virtual
effects appear ($\Lambda_{\rm LH}$) cannot be the same as the scale at
which the ultraviolet behavior of the Higgs mass is modified ($\Lambda_{\rm 
SM}$). This has forced us to abandon old versions of electroweak-breaking 
theories
with strong dynamics. From a theoretical perspective, extra dimensions
have brought  new tools into the game and there has been a great urge to
apply them at the weak scale. However, extra-dimensional theories are
non-renormalizable, and unitarity is violated at an energy of typically
few times the mass of the first KK mode. This implies that the cut-off
is not far from the electroweak scale and it opens the urgent question of
finding acceptable ultraviolet completions. The tools from extra dimensions 
have also given new hope for constructing theories of composite Higgs
bosons and to better understand their properties.      

Of course, in both desert and non-desert scenarios, 
many fundamental questions are postponed to 
physics at the cut-off, be it $M_{\rm Pl}$, 10~TeV, or 1~TeV. Although
theoretically this is equally unsatisfactory, the value of the cut-off 
makes a big difference
from a phenomenological point of view. When I discussed the little
hierarchy problem, I restricted myself to operators that satisfy 
the same symmetries as the SM Lagrangian. Had I introduced operators that
arbitrarily violate flavor, CP, lepton or 
baryon number, I would have obtained much more stringent
bounds on the corresponding mass scale. These bounds pose 
severe constraints on the dynamics of an ultraviolet 
completion at 10~TeV, while they are irrelevant for physics at $M_{\rm Pl}$. 

Finally, I want to point out that the question of desert versus non-desert
has important implications for the strategy of future collider planning.
In a scenario like conventional low-energy supersymmetry, a multi-TeV
linear collider seems the most appropriate next step after the LHC, because
precise studies of new-particle masses and properties are going to be
of the utmost importance. On the other hand, in non-desert 
scenarios, the new physics to be
discovered at
the LHC is just the first shell of a more complicated structure. Moving 
towards the highest possible energy then becomes a priority,
motivating research for hadron colliders in the 100--200~TeV region, like
the VLHC.  

\section*{Acknowledgements}
I wish to thank my colleagues at CERN for various discussions and, in 
particular, R.~Rattazzi for useful comments and A.~Strumia for producing
Fig.~\ref{fig1}.

\balance

\end{document}